\documentclass[preprint,prd,nofootinbib,tightenlines,amsmath,amssymb]{revtex4}
\usepackage{enumerate}
\usepackage{multirow}
\usepackage[utf8]{inputenc}
\usepackage{anysize}
\usepackage{textcomp}
\usepackage{epsfig}
\usepackage{graphics,color}
\usepackage{verbatim}
\usepackage{afterpage}
\usepackage{amsmath}    
\usepackage{soul}
\usepackage{xcolor,cancel}
\DeclareUnicodeCharacter{00A0}{ }

\usepackage{blindtext}

\catcode`\@=11
\def\sla@#1#2#3#4#5{{%
 \setbox\z@\hbox{$\m@th#4#5$}%
 \setbox\tw@\hbox{$\m@th#4#1$}%
 \dimen4\wd\ifdim\wd\z@<\wd\tw@\tw@\else\z@\fi
 \dimen@\ht\tw@
 \advance\dimen@-\dp\tw@ \advance\dimen@-\ht\z@
 \advance\dimen@\dp\z@
 \divide\dimen@\tw@ \advance\dimen@-#3\ht\tw@
 \advance\dimen@-#3\dp\tw@ \dimen@ii#2\wd\z@
 \raise-\dimen@\hbox to\dimen4{%
 \hss\kern\dimen@ii\box\tw@\kern-\dimen@ii\hss}%
 \llap{\hbox to\dimen4{\hss\box\z@\hss}}}}

\def\cpto{\mathrel {\vcenter {\baselineskip 0pt \kern 0pt
    \hbox{$H_{r.f.}$} \kern 0pt \hbox{$\longrightarrow$} }}}
\def\slashed#1{%
 \expandafter\ifx\csname sla@\string#1\endcsname\relax
{\mathpalette{\sla@/00}{#1}}
\fi}
\def\declareslashed#1#2#3#4#5{%
 \expandafter\def\csname sla@\string#5\endcsname{%
#1{\mathpalette{\sla@{#2}{#3}{#4}}{#5}}}}
 \catcode`\@=12
\declareslashed{}{/}{.08}{0}{D}
 \declareslashed{}{/}{.1}{0}{A}
 \declareslashed{}{/}{0}{-.05}{k}
 \declareslashed{}{/}{.1}{0}{\partial}
 \declareslashed{}{\not}{-.6}{0}{f}

\def\lsim{\mathrel {\vcenter {\baselineskip 0pt \kern 0pt
    \hbox{$<$} \kern 0pt \hbox{$\sim$} }}}
\def\gsim{\mathrel {\vcenter {\baselineskip 0pt \kern 0pt
    \hbox{$>$} \kern 0pt \hbox{$\sim$} }}}

\newcommand{\bea}{\begin{eqnarray}}
\newcommand{\eea}{\end{eqnarray}}

\begin{document}

\baselineskip=15pt
\preprint{}

\title{Realistic model for a fifth force
explaining\\ anomaly in ${^8Be^*} \to  {^8Be} \;{e^+e^-}$ Decay}

\author{Pei-Hong Gu$^{1}$\footnote{peihong.gu@sjtu.edu.cn}, Xiao-Gang He$^{1,2,3}$\footnote{hexg@phys.ntu.edu.tw}, }
\affiliation{
$^{1}$INPAC, Department of Physics and Astronomy, Shanghai Jiao Tong University, Shanghai, 200240\\
$^{2}$Department of Physics, National Taiwan University, Taipei, 10617\\
$^{3}$Physics Division, National Center for Theoretical Sciences, Hsinchu, 30013\\}

\date{\today}

\vskip 1cm
\begin{abstract}
A $6.8\,\sigma$ anomaly has been reported in the opening angle and invariant mass distributions of $e^+e^-$ pairs produced in ${^8Be}$ nuclear transitions. It has been shown that a protophobic fifth force mediated by a $17\,\textrm{MeV}$ gauge boson $X$ with pure vector current interactions can explain the data through the decay of an excited state to the ground state, ${^8Be^*} \to {^8Be}\, X$, and then the followed saturating decay $X \to e^+e^-$. In this work we propose a renormalizable model to realize this fifth force. Although axial-vector current interactions also exist in our model, their contributions cancel out in the iso-scalar interaction for ${^8Be^*} \to {^8Be} \,X$. Within the allowed parameter space, this model can alleviate the $(g-2)_\mu$ anomaly problem and can be probed by the LHCb experiment. Several other implications are discussed.

\end{abstract}

\pacs{PACS numbers: }

\maketitle

\section{Introduction}

Recently, studies of decays of an excited state of ${^8Be}$ to its ground state have found a $6.8\,\sigma$ anomaly in the opening angle and invariant mass distribution of $e^+e^-$ pairs produced in these transitions \cite{be-exp}. The discrepancy from expectations may be explained by unknown nuclear reactions or unidentified experimental effects, the observed distribution fits well by postulating the existence of a fifth force mediated by a new boson $X$ that is produced on-shell in ${^8Be^*} \to {^8Be}\; X$ and decays promptly via $X\to e^+e^-$. The authors of Ref. \cite{be-exp} have simulated this process, including the detector energy resolution, which broadens the $m_{ee}$ peak significantly. They find that the $X$ boson mass should be $m_X^{} = 16.7\pm0.35 (\textrm{stat})\pm0.5 (\textrm{sys})$ MeV.

It has been argued that the $X$ boson is likely a vector boson which couples non-chirally to the SM fermions \cite{jfeng},
\begin{eqnarray}
\label{eff}
L &=& -{1\over 4} X_{\mu\nu} X^{\mu\nu} + {1\over 2} m^2_X X_\mu X^\mu -X_\mu J^\mu_X\nonumber\\
&&\textrm{with}~~J_\mu^{}=\sum_{f=u,d,e,\nu_e^{},...}^{}e\varepsilon_f^{v}J_\mu^f=\sum_{f=u,d,e,\nu_e^{},...}^{}e\varepsilon_f^{v}\bar{f}\gamma_\mu^{} f\,.
\end{eqnarray}
Here the superscript “$v$" on $\varepsilon_f^{v}$ indicates the vector current coupling nature.

To explain the experimental data, the couplings $\varepsilon_f^v$ are determined from the following considerations. Assuming ${^8Be^*} \to {^8B_e}\;X$ followed by $X\to e^+ e^-$ saturating $X$ decay, one obtains \cite{jfeng}
\begin{eqnarray}
|\varepsilon^v_p + \varepsilon^v_n |\approx 0.011\,,~~|\varepsilon_e^{v}|\gtrsim 1.3\times 10^{-5}\,.
\end{eqnarray}
Here the fact that the interaction matrix element of $X$ with ${^8Be}$ and ${^8Be^*}$ is iso-scalar interaction has been taken into account which implies that the interaction is proportional to $\varepsilon_p^v + \varepsilon_n^v$. Note that
$\varepsilon_p^v = 2 \varepsilon_u^v + \varepsilon_d^v$ and $\varepsilon_n^v = \varepsilon_u^v + 2\varepsilon_d^v$.

The parameters are also constrained from other experimental data. An important one comes from $\pi^0 \to X \gamma$ where the decay width is proportional to $N_\pi =(Q_u \varepsilon^v_u - Q_d \varepsilon^v_d)^2$ resulting from a calculation similar to anomaly induced $\pi^0 \to \gamma\gamma$. Saturating the experimental limit $N_\pi= (2\varepsilon^v_u +\varepsilon^v_d)/9= \varepsilon^v_p/9 < \varepsilon^2_{max}/9$ with $\varepsilon_{max} = 8\times 10^{-4}$ \cite{pi0-exp}, one gets \cite{jfeng}
\begin{eqnarray}
-0.067 < {\varepsilon^v_p\over \varepsilon^v_n} < 0.078\;.
\end{eqnarray}
Therefore the coupling $\varepsilon_p^v$ is suppressed compared with $\varepsilon^v_n$. It has been suggested in Ref. \cite{jfeng} that the interaction might be protophobic with $\varepsilon_p^v = 0$. In this case,
\begin{eqnarray}
&&\varepsilon^v_u = \pm 3.7\times 10^{-3}\;,\;\;\;\;\varepsilon^v_d = \mp 7.4\times 10^{-3}\;. \label{data-q}
\end{eqnarray}
Requiring $\varepsilon^v_e$ to satisfy the lower bound from SLAC E141 experiment\cite{e141}, the stringent constraint from electron anomalous magnetic dipole moment $(g-2)_e$ \cite{pdg}, and also the precision studies of
$\bar{\nu}_e-e$ scattering from TEXONO \cite{texono}, one yields \cite{jfeng}
\begin{eqnarray}
&&2 \times 10^{-4} < |\varepsilon^v_e | < 1.4\times 10^{-3}\;,\;\;\;\;|\varepsilon^v_e \varepsilon^v_{\nu_e}|^{1/2}< 7\times 10^{-5}\;. \label{data-e}
\end{eqnarray}

In general the $X$ boson may also have axial-vector current couplings to the SM fermions, i.e.
\begin{eqnarray}
\label{axial}
e \varepsilon^a_f \bar f \gamma_\mu \gamma_5f\,.
\end{eqnarray}
However, the $X$ boson is only allowed to give an extremely tiny contribution to the decay width of $\pi^0_{} \rightarrow e^{+}_{}e^{-}_{}$. This sentences the case that the $X$ boson has a sizable axial-vector current interaction with both the electron and the first-generation quarks. Alternatively, the $X$ boson can be allowed to sizably couple to the axial-vector current of either the electron or the first-generation quarks.

The $X$ boson interactions discussed above are based on an effective theory approach. It would be interesting to have the $X$ boson be part of a consistent theory respecting the standard model (SM) symmetry $SU(3)^{}_C \times SU(2)^{}_L \times  U(1)^{}_Y$. For this purpose, one must consider more constraints from both theoretical and experimental constraints which make the task non-trivial. In this letter we show the first successful realization of this goal. Specifically we consider new gauge symmetries $U(1)_{Y'}$ and $U(1)_X$ in addition to the SM gauge group. The $U(1)_{Y'}$ gauges certain variations of generation number difference without beyond the SM fermions. The $X$ boson is just the $U(1)_X$ gauge boson and couples to the SM fermions through the $U(1)_{Y'}$ and $U(1)_X$ kinetic mixing. In our model, the $X$ boson has both vector and axial-vector current couplings. The vector current interactions are protophobic, while the axial-vector currents are not protophobic but have no contributions to the iso-scalar interaction for ${^8Be^*} \to {^8Be}\; X$. Within the allowed parameter space, this model can alleviate the $(g-2)_\mu$ anomaly problem and can be tested by the LHCb experiment.

\section{A realistic model}

From theoretical side, the SM fermions appear in form of chiral fields, implying that introduction of new gauge boson interaction may generate gauge anomalies which is not allowed. We here consider to construct a model free of gauge anomaly by using the anomaly cancellation among different generations, similar to the gauge anomaly free model for $L_i - L_j$ in the literature \cite{a-free}. The appearance of chiral fields in general makes the $X$ boson interaction not purely vector current type which may lead to complications and need to be carefully treated. Also the $X$ boson may interact with different generations in general, there are more constraints from data. It is remarkable that our model can explain all of the data well. We provide the details in the following.

The key to our construction is to have a protophobic vector current first and then accommodate the constraints from ${^8Be^*}\to {^8Be} X$ and $\pi^0 \to X \gamma$. To achieve this we introduce a $U(1)_{Y'}$ gauge symmetry whose vector current is protophobic.
The assignment of quantum numbers for the three generations of fermions, under the $SU(3)_C\times SU(2)_L\times U(1)_Y\times U(1)_{Y'}$, are as the following
\begin{eqnarray}
&&Q^1_L: (3,\;2,\;1/6)(-1)\;,\;\;\;\;\;\;u^1_R: (3,\;1,\;2/3)(5)\;,\;\;\;\;\;\;\;\;d^1_R: (3,\;1,\;-1/3)(-7)\;,\nonumber\\
&&L^1_L: (1,\;2,\;-1/2)(\beta)\;,\;\;\;\;\;\;\;e^1_R: (1,\;1,\;-1)(\beta)\;,\nonumber\\
&&Q^2_L: (3,\;2,\;1/6)(1)\;,\;\;\;\;\;\;\;\;\;u^2_R: (3,\;1,\;2/3)(-5)\;,\;\;\;\;\;d^2_R: (3,\;1,\;-1/3)(7)\;,\nonumber\\
&&L^2_L: (1,\;2,\;-1/2)(-\beta)\;,\;\;\;\;e^2_R: (1,\;1,\;-1)(-\beta)\;,
\end{eqnarray}
and the third generation does not have any $U(1)_{Y'}$ charges. One can easily check that the model is free of gauge anomaly. Expanding the interactions between the $U(1)_{Y'}$ gauge boson $Y'$ and the SM fermions, $Y'^{}_\mu J^\mu_{Y'}$, we have the current coupling to the $Y'$ field,
\begin{eqnarray}
J^\mu_{Y'} &=& g_{Y'}^{} \left [\bar u\gamma^\mu (4 + 6\gamma_5) u - \bar d \gamma^\mu(8 +6 \gamma_5) d +\beta \bar e \gamma^\mu e +{\beta \over 2} \bar \nu_e \gamma^\mu(1-\gamma_5)\nu_e\right]\nonumber\\
&-&g_{Y'}^{} \left [\bar c\gamma^\mu (4 + 6\gamma_5) c - \bar s \gamma^\mu(8 +6 \gamma_5) s +\beta \bar \mu \gamma^\mu \mu +{\beta \over 2} \bar \nu_\mu \gamma^\mu(1-\gamma_5)\nu_\mu\right]\;, \label{j}
\end{eqnarray}
with $g_{Y'}^{}$ being the $U(1)_{Y'}$ gauge coupling.

We cannot identify the $Y'$ boson as the desired $X$ boson. The reason is that the Higgs scalars giving the SM fermion masses will have non-trivial quantum numbers for both the SM and $U(1)_{Y'}$ gauge groups and
will contribute to the $W$, $Z$ and $Y'$ gauge boson masses. If $Y'$ is $X$, it must have a small mass $17\,\textrm{MeV}$, and the involved vacuum expectation values (VEVs) should be much smaller than the electroweak scale for a $g_{Y'}^{}$ explaining the anomalous ${^8Be^*} \to {^8Be}\; {e^+e^-}$. Another problem is that the couplings of neutrinos to $Y'$ are too large to satisfy the constraints mentioned previously. These problems must be solved for a realistic model.

We solve the light mass problem by introducing an additional gauge symmetry $U(1)_X$, under which the SM fermions are trivial. But through a  kinetic mixing of $U(1)$ gauge fields \cite{u1-mixing}, $-(\epsilon/2) Y'_{\mu\nu} X^{\mu\nu}$, the $X$ boson does interact with the SM fermions. By diagonalizing and normalizing the gauge fields $Y'$ and $X$ properly, up to the leading order in $\epsilon$, we give the couplings of the $X$ boson to the $J^\mu_{Y'}$ current,
\begin{eqnarray}
\epsilon X_\mu J^\mu_{Y'}\;.\label{x}
\end{eqnarray}

Since the $X$ boson does not carry the SM gauge group quantum numbers, one can generate a small mass $m^2_X = (g_X x_\rho v_{\rho})^2$ by introducing an SM-singlet scalar $\rho$ with a $U(1)_X$ charge $x_\rho$ and a VEV $v_{\rho}/\sqrt{2}$. Here $g_X^{}$ is the $U(1)_X$ gauge coupling. Clearly, this small mass $m_X^{}$ does not affect the usual electroweak scale. We will assume the $X$ boson to have a mass of $17\,\textrm{MeV}$. At the same time, one can introduce another scalar singlet $\sigma$ with a $U(1)_{Y'}$ charge $y'^{}_{\sigma}$ and a VEV $v_{\sigma}/\sqrt{2}$ to contribute to the $Y'$ mass with
$m^2_{Y'} = (g_{Y'} y'_{\sigma} v_{\sigma})^2$. Assuming $m_{Y'}$ of the order of $\textrm{TeV}$, the contributions from the Higgs scalars transforming as the SM iso-doublets can be neglected since their VEVs are at the electroweak scale.

We now discuss how to suppress the couplings of the $X$ boson to the electron neutrino $\nu_{e}^{}$. This is achieved by mixing $\nu_{e}^{}$ with a new vector-like fermion $S=S_L^{}+S_R^{}$ which is a singlet under the SM and $U(1)_{Y'}$ gauge groups but carry a $U(1)_X$ charge $x_{S}$. One can also introduce
three gauge-singlet fermions $N_{Ri}~(i=1,2,3)$ to facilitate a canonical seesaw mechanism for generating the small neutrino masses. Let us take the first generation into account for illustration. With three iso-doublet Higgs scalars $\phi_e(0,0)$, $\phi_{\nu_e}(\beta,0)$, and $\eta(\beta, -x_{S})$, where the brackets following the fields describe the transformations under the $U(1)'^{}_Y\times U(1)_X^{}$ gauge groups, the terms responsible for the first-generation lepton masses are
\begin{eqnarray}
\label{firstg}
L = - y^{}_e \bar L^1_L L\tilde{\phi_e} e^{}_R-y^{}_N \bar L^1_L\phi_{\nu_e} N^{}_R-\frac{1}{2}M_N \bar{N}_R^c N_R^{}- f_S^{}\bar L^1_L\eta S_R^{} - m_S^{}\bar{S}_L^{} S_R^{}+\textrm{H.c.}\,.
\end{eqnarray}

We emphasize that the mixing between the vector-like fermion and the electron neutrino will not be stringently constrained by the neutrino masses, instead, it will affect the Dirac equations of the left-handed electron neutrino, i.e.
\begin{eqnarray}
\mathcal{L}\supset i\bar{\nu}^{}_{Le}\left(1+U_S^{}\right)\partial\!\!\!/\, \nu_{Le}^{}-\frac{1}{2}\bar{\nu}_{Le}^{}m_\nu^{}\nu_{Le}^{c}
+\textrm{H.c.}\,.
\end{eqnarray}
Here $U_S^{}$ is a real number mediated by the vector-like fermion while $m_\nu^{}$ is the neutrino mass suppressed by the right-handed neutrinos,
\begin{eqnarray}
U_S^{}=f_S^{}\frac{v_\eta^2}{2m_S^{2}}f_S^{\dagger}\,,~~m_\nu^{}=-y_N^{}\frac{v_{\phi_e}^2}{2M_N^{}}y_N^{T}\,.
\end{eqnarray}
We then should normalize the left-handed electron neutrino and its mass by
\begin{eqnarray}
\label{nor}
&&(1+U_S^{})^{\frac{1}{2}}_{}\nu_{Le}^{} \rightarrow \nu_{Le}^{}\,,\;\;\;\;(1+U_S^{})^{-\frac{1}{2}}_{}m_\nu^{}(1+U_S^{T})^{-\frac{1}{2}}_{}\rightarrow m_\nu^{} \,.
\end{eqnarray}
In principle, the right-handed neutrinos will also modify the kinetic term of the left-handed electron neutrino. However, this contribution is of the order of $m_\nu^{}/M_N^{}$ and hence is negligible.

By integrating out the vector-like fermion, a term of $-x_S^{}g_X^{}\bar{\nu}_{Le}^{}U_S^{}\gamma_\mu^{}\nu_{Le}^{}X^\mu_{}$ will be generated. Including the normalization according to Eq. (\ref{nor}), one finds the effective coupling of the electron neutrino $\nu_e^{}$ to the $X$ boson should be
\begin{eqnarray}
{\beta\epsilon g_{Y'}^{} \over 2} \bar{\nu}_{e}^{}\gamma^{}_\mu (1-\gamma_5) \nu_{e}^{} X^\mu_{}\to  {\beta \epsilon g_{Y'}^{}\over 2}  \frac{1-\frac{g_X^{}x_S^{}}{\beta\epsilon g_{Y'}}U_{S}^{}}{1+U_{S}^{}}\bar{\nu}_{e}^{}\gamma^{}_\mu (1-\gamma_5) \nu_{e}^{} X^\mu_{}\,.\label{nu}
\end{eqnarray}
With an appropriate choice of parameters, the coupling of $X$ to $\nu_e$ can be supressed, even to zero if $g_X^{}x_S^{} U_{S}^{} = \epsilon g_{Y'}^{} \beta$. The demonstrations in Eqs. (\ref{firstg}-\ref{nu}) can be generalized for all of the three generations by introducing more iso-doublet Higgs scalars and vector-like fermions with proper $U(1)'^{}_Y\times U(1)_X^{}$ charges. In this case, the numbers $U_S^{}$ and $m_\nu^{}$ should be understood as a hermitian matrix and a symmetric matrix, respectively.

\section{The fifth force}

Combining Eqs.(\ref{j}), (\ref{x}) and (\ref{nu}), we derive the parameters $\varepsilon_f^v$ in the effective theory (\ref{eff}) by
\begin{eqnarray}
\varepsilon^v_u =-\frac{4 \epsilon g_{Y'}^{}}{e}\,,~~\varepsilon^v_d = \frac{8 \epsilon g_{Y'}^{}}{e}\,,~~
\varepsilon^v_e =  - \frac{ \beta\epsilon g_{Y'}^{}}{e}\,,~~ \varepsilon^v_{\nu_e} =  {\beta\epsilon  g_{Y'}^{} \over 2 e} {1-g_XU_{S}/(\beta\epsilon g_{Y'})\over 1+ U_{S}}\,.~~
\end{eqnarray}
Obviously, $\varepsilon_p^v = 2 \varepsilon_u^v + \varepsilon^v_d = 0$. So, the vector current interaction of the $X$ boson is protophobic type as proposed in Ref. \cite{jfeng}. At this moment, one may have naively concluded that they can easily fit the required numbers for explaining the ${^8Be^*} \to {^8Be}\; e^+e^-$ data as given in Eqs. (\ref{data-q}). However, the model above also contains axial-vector current interactions (\ref{axial}) with
\begin{eqnarray}
\varepsilon^a_u= - \frac{6\epsilon g_{Y'}^{}}{e}\,,~~\varepsilon^a_d =\frac{6 \epsilon g_{Y'}^{}}{e}\,,~~\varepsilon_e^a =  0\,,~~\varepsilon^a_{\nu_e}=-{\beta \epsilon  g_{Y'}^{}\over 2 e} {1-g_XU_{S}/(\beta\epsilon g_{Y'})\over 1+ U_{S}}\,.
\end{eqnarray}
Therefore the model is actually protophobic only in the vector current interactions. It is necessary to check if the axial-vector current interactions can satisfy the experimental data. Remarkably, the axial-vector current interactions between the $X$ boson and the $u$ and $d$ quarks are proportional to $\varepsilon^a_u \bar u \gamma^\mu \gamma_5 u + \varepsilon_d^a \bar d \gamma^\mu \gamma_5 d$ which now is the sum of a zero iso-scalar current $[(\varepsilon^a_u + \varepsilon_d^a)/2] (\bar u \gamma^\mu \gamma_5 u +
\bar d \gamma^\mu \gamma_5 d)\equiv 0$ and a nonzero iso-vector current $[(\varepsilon^a_u - \varepsilon_d^a)/2] (\bar u \gamma^\mu \gamma_5 u -
\bar d \gamma^\mu \gamma_5 d)\equiv\!\!\!\!\!\!/ \,\,0$ \cite{isoscalar}. On the other hand, the observed ${^8Be^*} \to {^8Be}\, X$ process is irrelevant to the iso-vector currents because the initial and final hadrons are both isospin singlets. The iso-vector interaction may induce some physical effects, such as $\pi^0 \to e^+e^-$. However, the electron only has a vector current interaction with the $X$ boson so that the contribution from our model to $\pi^0\to e^+e^-$ can be identically zero. No constraint can be obtained from this consideration.

The protophobic nature in the vector current interactions of the $X$ boson results in an interaction term $\pi^0 \tilde X_{\mu\nu} F^{\mu\nu}$ through triangle anomaly diagram which generates $\pi^0 \to \gamma\gamma$ decay. Here $F^{\mu\nu}$ is the photon field strength. Experimental limit on search of $\pi^0 \to X \gamma$ constrains the interaction to be protophobic as mentioned earlier. Our model contains the axial-vector current couplings of the $X$ boson to the $u$ and $d$ quarks. Naively, one would expect the emergence of an interaction term of the type of $\pi^0 X_{\mu\nu}F^{\mu\nu}$ which affects the result. However, this term does not appear since it violates CP and is therefore forbidden. The analysis of $\pi^0 \to X \gamma$ in Ref. \cite{jfeng} still hold in our model.

To explain the observed anomalous $^{8}_{}\!Be$ nuclear transitions and fulfill all of the other experimental limits, one needs
$|\varepsilon_n^v| = |12\epsilon g_{Y'}/e| = 0.011$ and $|\varepsilon_e^v| =|\beta\epsilon g_{Y'}/e| <1.4\times 10^{-3}_{}$ which result in
\begin{eqnarray}
\epsilon g_{Y'} = 2.78\times 10^{-4}\,,~~|\beta|<1.53\,.
\end{eqnarray}

\section{Other implications}

Since in our model, the first two generations of charged fermions couple to the $X$ boson with a same strength, in particular, $\varepsilon_d^{v} = \varepsilon_s^{v} =2\varepsilon_n^{v}/3$, there may be constraints from  data on $X$ production from other quarks. The value for $|\varepsilon_s^{v}| = 0.0073$ is
at tension with the  boundary of the $90\%$ c.l. allowed region from KOEL data\cite{koel} on $\phi \to \eta X$. But allowed at 3$\sigma$ c.l.. Improved data can test the model further.

Furthermore, we have $|\varepsilon_\mu^{v} |= |\varepsilon_e^{v}|$, which has an effect on $(g-2)_\mu^{}$. One can calculate the $X$ boson contribution to $\Delta a_\mu$ which
has a $3\,\sigma$ deviation, $\Delta a_\mu^{} = 288(80)\times 10^{-11}_{}$ \cite{aoyama2012}.
Using the $3\,\sigma$ upper bound of $\varepsilon_e^{v}=1.4\times 10^{-3}_{}$ ($\beta = 1.53$), we obtain $\Delta a_\mu = 152\times 10^{-11}$ from the $X$ boson contribution which improves the deviation to $1.5\,\sigma$.

We now discuss possible ways to further test the model. Besides continuing similar experiments with higher sensitivity for those already provided constraints, it would be good to find new ways for testing the model. One may carry out $e^+_{} e^-_{}\to \gamma X$ followed by measuring $e^+_{}e^-_{}$ with a center of mass energy $\sqrt{s}$ at BES III and also at BELLE II. Since in our model $\varepsilon_e^{v}$ is constrained to be less than $1.4 \times 10^{-3}_{}$, the cross section is typically  less than $ 10^{-2}_{}\,\textrm{fb}$ which may be too small to be measured experimentally in the near future. At hadron collider because $|\varepsilon_{d}^{v}|$ is as large as $7\times 10^{-3}_{}$, the cross section for, $pp \to \gamma X+\textrm{jets}$, may be larger. However, in the hadronic back ground the measurement will be very challenging.

Exclusive decay of a meson $A$ to $B X$ followed by measuring $e^+_{}e^-_{}$ from on-shell $X$ decay may be very hopeful. If the initial state $A$ is a state with two constituent quarks (a quark with an anti-quark have the same absolute electric charge $|Q_q^{}|$), one then obtains, for the vector part of the current interaction. $R(X/\gamma, Q_q)=
{Br(A\to B X)_{Q_q}/Br(A\to B\gamma)_{Q_q}} = (\varepsilon_q^v/Q_q)^2 $.
Assuming $X\to e^+e^-$ saturating the $X$ decay, for $Q_q = 2/3$ and $Q_q = -1/3$, we have, respectively
$R(X/\gamma,\;2/3)= \left|\frac{1}{2}\varepsilon_n^{v}\right|^2_{}\approx 3.0\times 10^{-5}$ and
$R(X/\gamma,\;-1/3) = \left|2\varepsilon_n^{v}\right|^2_{}\approx 4.8\times 10^{-4}$. When the axial-vector current contributions are included which will add terms proportional to $|\varepsilon^a_q|^2$, ratios become larger. So the numbers $3.0\times 10^{-5}$ and $4.8\times 10^{-4}$ represent lower bounds for the ratios.

The above bounds can be used to study radiative ($X$ boson) decays of the vector mesons $J/\psi$ or the flavored vector mesons $D^{*0}$ into a spin zero meson. We find the most promising decay mode is $D^{*0} \to D^0 X\to D^0 e^+e^-$ for the reasons that $D^{*0}\to D^0 \gamma$ has a large branching ratio $(38.1\pm 2.9)\% $ \cite{pdg} and a large number of this decay can be copiously produced and studied at the LHCb. At the LHC run III, the LHCb may have an integrated luminosity of $15\,\textrm{fb}^{-1}$ which means that the event number for $D^{*0} \to D^0 \gamma$ can reach about $5\times 10^{12}_{}$. The analysis for constraining $\varepsilon_c^{v}=-\varepsilon_n^{v}/3$ is similar to that carried out for constraining the dark photon mixing parameter in Ref. \cite{wx} where it was shown that for $m_X^{}\simeq 17\,\textrm{MeV}$ the LHCb sensitivity for the mixing parameter can reach about
$2.4\times 10^{-5}$ with an integrated luminosity of $15\,\textrm{fb}^{-1}_{}$. Normalizing their notation to ours, the sensitivity for
$|\varepsilon_c^{v}|$ can be $1.6\times 10^{-5}$. Our model can be tested at the LHCb.

In the above discussions, the third-generation fermions do not interact with the $X$ boson. However, it may turns out that
the third-generation quarks interact with the $X$ boson, but the second-generation does not. In this case,
the above formulae can be used to study radiative ($X$ boson) decays of the vector mesons $\Upsilon$ into a spin zero meson, or
radiative ($X$ boson) decays of the flavored vector mesons $B^{*0}_d$ and $B^{*0}_s$ into a spin zero meson.

\section{Conclusions}

In summary, we have proposed a realistic gauge anomaly free model with a $17\,\textrm{MeV}$ $X$ gauge boson mediating a fifth force to explain the anomaly reported in ${^8Be^*} \to {^8Be}\; e^+e^-$. In our model, the $X$ boson has both of the vector and axial-vector current couplings to the SM fermions. The vector current interactions have a protophobic nature. Meanwhile, the contribution from the axial-vector currents cancels out in the iso-scalar interactions for ${^8Be^*} \to {^8Be}\; X$. Furthermore, the model allows us to suppress the unexpected couplings of the $X$ boson to the electron neutrino. Within the allowed parameter space, the model can alleviate the anomaly in $(g-2)_\mu$. The $X$ boson also couples to the second or third generation of quarks and hence may induce $D^{*0} \to D^0 X \to D^0\;{e^+e^-}$ or $B^{*0}_{d,s} \to B^0_{d,s} X \to B^0_{d,s}\;{e^+e^-}$ which can be studied at the LHCb to probe the parameter space for explaining ${^8Be^*} \to Be X \to Be\; {e^+e^-}$. For generating the required fermion masses, we need introduce multi iso-doublet Higgs scalars carrying different $U(1)_{Y'}$ and/or $U(1)_X$ charges. This means rich flavor changing phenomena including the anomaly in $h \to \mu\tau$ from the LHC and the anomalies in $b\to s \mu^+\mu^-$ transitions shown in experimental data. We will present detailed studies elsewhere.

\begin{acknowledgments}

PHG was supported by the Shanghai Jiao Tong University (Grant No. WF220407201) and the Recruitment Program for Young Professionals (Grant No. 15Z127060004). XGH was supported in part by MOE Academic Excellent Program (Grant No.~102R891505) and MOST of ROC (Grant No.~MOST104-2112-M-002-015-MY3), and in part by NSFC (Grant Nos.~11175115 and 11575111) of PRC. This work was also supported by the Shanghai Laboratory for Particle Physics and Cosmology (Grant No. 11DZ2260700).

\end{acknowledgments}

\end{document}